\documentclass[prl,showpacs,amsfonts,amssymb,amsmath,twocolumn]{revtex4}
\usepackage{times}
\usepackage{graphicx}

\begin{document}

\title{Casimir-like force arising from quantum fluctuations in a slow-moving dilute Bose-Einstein condensate}

\author{D. C. Roberts and Y. Pomeau} \affiliation{Laboratoire de
  Physique Statistique de l'Ecole normale sup\'erieure, Paris, France} 
\begin{abstract}
We calculate a force due to zero-temperature quantum fluctuations on a stationary object in a moving superfluid flow.  We model the object by a localized potential varying only in the flow direction and model the flow by a three-dimensional weakly interacting Bose-Einstein condensate at zero temperature.  We show that this force exists for any arbitrarily small flow velocity and discuss the implications for the stability of superfluid flow.
\end{abstract}

\maketitle

Although there are various definitions of superfluidity \cite{leggett}, one 
of
the defining features of a superfluid is the existence of a critical velocity 
below which the superfluid flows without dissipation.  Landau argued that, by 
performing a Galilean transformation on the ground state of a uniform 
superfluid, the superfluid would become unstable above a well-defined critical velocity due to the creation of 
quasiparticles \cite{landau}.  If one assumes that it is only through the creation of quasiparticles that
dissipation can occur at $T=0$, then one 
can infer that a stationary object in a slow-moving superfluid  (with  a flow 
velocity well below the critical velocity), would remain in a metastable stationary state as there would be no force acting on 
this object.

In this letter, we show that the phenomenological picture of superfluid flow - namely the existence of a metastable state below a critical velocity - is incomplete and problematic.  We illustrate this by examining the case of a localized potential fixed in the flow of a three dimensional dilute Bose-Einstein condensate.
 Specifically, we show that a
force arises from the scattering of zero-temperature quantum fluctuations, 
an Landau ignored in his argument for a critical velocity.  We demonstrate the existence of this force in an infinitely extended condensate at all nonzero flow velocities (including velocities much lower than Landau's critical).  

Casimir \cite{casimir} first showed that zero-temperature quantum fluctuations in an electromagnetic (EM) 
vacuum give rise to
an attractive force between two closely spaced perfectly conducting plates.  A Casimir-like force, 
$F_{BEC}$, can be shown to arise from the zero-point quantum fluctuations in a dilute BEC, where infinitely thin and infinitely repulsive plates immersed in a zero-temperature three-dimensional dilute BEC replace Casimir's perfect conducting plates.  $F_{BEC}$ is given by (to leading order) 
\begin{equation}
F_{BEC} \approx -\frac{\pi^2}{480} \frac{\hbar c_s \sigma}{d^4} 
\end{equation}
where $c_s$ is the speed of sound in the dilute BEC, $d$ is the distance between the plates, and $\sigma$ is the area of the plates. Note 
that the Casimir and Casimir-like forces in the EM vacuum and the BEC vacuum, respectively,
have the same expression except that the speed of sound replaces the speed of 
light in the BEC case \cite{roberts}. (The expressions also differ by a factor of 2  due to the fact that  EM 
modes have two transverse polarizations whereas  phonons have only one polarization.)  Both of these forces arise because boundary conditions are imposed on the quantum fluctuations.

Similarly, we posit that a Casimir-drag force exists on an object in a moving dilute BEC.  No direct EM analogy can be drawn for this situation because no absolute rest frame (where the relative motion of an object can be measured) exists for an EM vacuum \cite{emdrag}.    Nevertheless we maintain that in a superfluid flow at zero temperature, modeled as a weakly interacting BEC,  around a stationary object (or, by a Galilean transformation, a moving object in a stationary
superfluid), a Casimir-like force should arise due to the boundary 
conditions on the quantum fluctuations of a BEC vacuum.  In the specific case of a weak potential varying only in the flow direction, we show that this drag force exists at all nonzero flow velocities, i.e.  the effective critical velocity, defined as the velocity below which the flow is dissipationless,  is zero for this system.  

We now calculate the drag force arising from these quantum fluctuations.  Momentum is not, in general, conserved in our system because the stationary object, which is described by the potential $\Phi(r)$, breaks the translational symmetry \cite{emcons}.  In general, a force on a moving object described by a potential $\Phi(r)$ can be written in second quantized notation at zero temperature as
\begin{equation}
\label{force}
{\vec F}=-\int d^3r \langle \hat \psi^\dag (r) [{\vec \nabla} \Phi(r)] \hat \psi (r) \rangle_{T=0},
\end{equation}
where $\hat \psi(r)$ and $\hat \psi^\dag(r)$ are field operators that describe the weakly interacting BEC flow and obey the standard boson commutation relations and the expectation value is taken at $T=0$.  $T=0$ is not well defined in the scattering problem discussed in this letter so one can view this simply as a convenient label of the quantum state that we define in detail below.

We model the superfluid as a weakly interacting three-dimensional
condensate
characterized by an interparticle contact pseudopotential, $g
\delta^{(3)}(r)$, where $g$ is determined by the 2-particle positive 
scattering
length $a_{sc}$ and the mass $m$ of the
atoms such that $g=4 \pi \hbar^2 a_{sc}/m$.  We assume the condensate to be dilute such 
that $\sqrt{\rho_0
a_{sc}^3}\ll1$ where $\rho_0$ is the condensate number density.

To calculate the force due to quantum fluctuations on a stationary object in a superfluid flow, we assume for
simplicity that the object is described by a weak symmetric potential 
that
varies only in the flow direction (which we take as the $x$-direction), i.e. $\Phi({\bf r})= \eta \Phi(x)$ where 
$\Phi(x)=\Phi(-x)$
and $\eta\ll1$ (The parameter $\eta$ gives the order of magnitude of the external potential.) This situation, which can, in principle, be realized in current dilute BEC experiments, is a specific case chosen to show the existence of a finite Casimir-like drag force at any arbitrarily small velocity.  
 We will place further restrictions on this potential in the course of this letter
as needed.

Because we consider a potential varying  only in the flow direction, the integrand in eq. (\ref{force}) is only a function of the positional variable $x$ (the $y$ and $z$ dependencies implicit in $\hat \psi (r)$ and $\hat \psi^\dag (r)$ cancel out), which allows the simplification of the force expression to
\begin{equation}
F_x=-A \eta \int^\infty_{-\infty} d x \langle \hat \psi^\dag (r) \frac{d \Phi(x)}{d x}  \hat \psi (r) \rangle_{T=0},
\end{equation}
where $A$ is the cross-sectional area of the object in the flow. 

Although the lack of translational symmetry (due to the presence of the object) makes the existence of a drag force possible, it does not imply that there will necessarily be a drag force.  For example, if the small quantum fluctuations are ignored, the bosonic field operator $\hat \psi$ can be approximated by the classical mean field $\Psi^{(0)}$ \cite{mac},  whose behavior is determined by  the Gross-Pitaevskii equation (GPE).  Working, as we will do throughout this letter unless specified otherwise, in dimensionless variables in which the length scale is normalized by the healing length given by $(8 \pi \rho_0 a_{sc})^{-1/2}$ and $\Psi$ is normalized by $\sqrt{\rho_0}$, the GPE can be written as 
\begin{equation}
\label{GP} (\hat T+\Phi(x)-\mu) \Psi^{(0)}(r)+| \Psi^{(0)}(r) |^2 \Psi^{(0)}(r)=0,
\end{equation}
where $\hat T \equiv - \vec{\nabla}^2+\sqrt{2} i q \frac{\partial}{\partial x}+q^2/2$, the dimensionless speed is given by $q= c/c_s$, $c$ is the speed of the flow at $x=\infty$, $c_s=\sqrt{\rho_0 g/m}$ is the speed of sound, and $\mu=1+q^2$ is the chemical potential (determined by imposing $\Psi^{(0)}(r)=1$ at $x=\infty$).  The mean field force arising from the potential $\Phi(x)$, given by $ -A \eta \int dx |\Psi^{(0)}(r)|^2 \frac{d \Phi(x)}{d x}$,  can be shown to be zero below a certain critical flow velocity.  This critical flow velocity (as measured far from the potential) in a nonuniform medium is always smaller than Landau's critical velocity in a uniform medium (which in the dilute gas is given by the speed of sound) due to nonlinear effects such as vortex shedding \cite{vc} or the creation of gray solitons \cite{Hakim}, and is the velocity at which the maximum local fluid velocity reaches the speed of sound \cite{modv}. 

%Since the potential is assumed to be real and symmetric, implying $\tilde 
%V(\lambda)=\tilde V(-\lambda)$ and $\tilde V^*(\lambda)=-V(\lambda)$ so 
%$F^{(0)}_{GPE}=0$, where the superscript signifies perturbation theory not 
%to be confused with $1/n_0$.

If we go beyond the mean field approximation and take into account quantum 
fluctuations,  the bosonic quantum field operator $\hat \psi$ can be split into a macroscopic 
classical field $\Psi^{(1)}$ and a small quantum fluctuation operator $ \hat 
\phi$: $\hat \psi=\Psi^{(1)} +\hat \phi$.  $\Psi^{(1)}$ is an improved approximation of the condensate wave function as compared to $\Psi^{(0)}$ because it includes the effects of the quantum fluctuations. 
Including the fluctuation operator in the analysis leads to a depletion of the ground state \cite{bog} and a correction to the ground state energy \cite{Lee}, both on the order of the diluteness parameter $\sqrt{\rho_0 a_{sc}^3}$, which must be small in order for the Bogoliubov theory in this paper to be valid.  
The effects of this quantum depletion and their correlations are well known (see references in \cite{roberts}).

The force due to the boundary conditions imposed by the potential on the quantum fluctuations is, in general, not zero (even below the critical flow velocity in a nonuniform system given by the GPE) and can be written as
\begin{equation}
\label{fdiv}
{F_x}=-A \eta \int dx (| \Psi^{(1)}(r)|^2+ \langle \hat \phi^\dag (r) \hat \phi 
(r) \rangle_{T=0})\frac{d \Phi(x)}{d x}. 
\end{equation}

The fluctuation operator can be expanded in terms of $\hat
\alpha_{k}$ and $\hat \alpha_{k}^\dag$ - the quasiparticle annihilation and creation operators, respectively -  such that $\hat \phi(r)=\sum_{k} \left( u_{k}(r) \hat
\alpha_{k} -v^*_{k}(r) \hat \alpha_{k}^\dag \right),$
where the sum excludes the condensate mode.  For our system of weakly interacting particles to be described by the non-interacting quasiparticles, the quasiparticle amplitudes, $u_{k}(r)$ and $v_{k}(r)$, must obey the Bogoliubov-de Gennes (BdG) equations  \cite{bog, fetter},
\begin{equation}
\hat {\cal L} u_{k}(r) -(\Psi^{(1)})^2 v_{k}(r) =
E_k u_{ k}(r)
\end{equation}
\begin{equation}
\hat {\cal L}^*  v_{k}(r) -(\Psi^{(1)*})^2 u_{k}(r)
= -E_k v_{k}(r),
\end{equation}
where $\hat {\cal L} = \hat T + \Phi(x) - \mu +2 |\Psi^{(1)}|^2$ and $k$ is the dimensionless momentum normalized by the healing length.  The energy eigenvalue for the moving BEC flow is $E_k=\sqrt{2}qk_x+E_B$ where the Bogoliubov dimensionless dispersion relation for a BEC at rest is given by $E_B=k \sqrt{k^2+2}$.

Since the BdG equations describe an effective scattering problem for the quasiparticle amplitudes, we can solve for $u_k(r)$ and $v_k(r)$ by specifying the incoming $u_k(r)$ and $v_k(r)$.  We impose a condition (see \cite{scatter} for details) on the incoming $u_k(r)$ and $v_k(r)$ in terms of measurable quantities far from the potential that fully determine the quantum state in this problem, i.e. the quantum state used in the expectation value in eq. (\ref{fdiv}). 
Despite the fact that the state at zero temperature is technically not well defined for this scattering problem, this label remains a convenient way to denote the quantum state relevant to eq. (\ref{fdiv}) because the state is annihilated by $\hat \alpha$.  
The expectation values taken with respect to this state (or at "$T=0$") can be written in terms of the quasiparticle amplitudes, i.e. $\langle \hat \phi^\dag (r) \hat \phi (r) \rangle_{T=0} = \sum_{k} |v_k(r)|^2$.

The condensate wave function modified by the quantum fluctuations is given by the generalized GPE (GGPE) \cite{Castin} 
\begin{eqnarray}
\label{GGPE}
 &&(\hat T+\Phi(x)-\mu) \Psi^{(1)}(r)+| \Psi^{(1)}(r) |^2 \Psi^{(1)}(r)-\chi(r)\Psi^{(1)}(r)+\nonumber \\
&&\sum_{k} \left[ 2 |v_k(r)|^2\Psi^{(1)}(r)-u_k(r)v^*_k(r)\Psi^{(1)}(r) \right]=0.
\end{eqnarray}
where the term proportional to $u_k(r)v^*_k(r)$ is ultraviolet divergent because of the contact potential approximation and must be renormalized (see references in \cite{morgan}).  The term $\chi(r) \Psi^{(1)}(r)$ ensures the orthogonality between the excited modes and the condensate \cite{morgan} and is given by $\chi(r)=\sum_{k} c_k v^*_k(r)$ where $c_k=\int d^3r  | \Psi^{(1)}(r) |^2(\Psi^{(1)*}(r) u_k(r)+ \Psi^{(1)}(r) v^*_k(r))$.  Because of the properties of $\Phi(x)$ (described below), neither $\chi(r) \Psi^{(1)}(r)$ nor the renormalization term contribute to the dominant order of the calculation below.  Because we assume the condensate to be dilute, the fluctuation terms are small (on the order of $\sqrt{\rho_0 a_{sc}^3}$) so the GGPE (eq. \ref{GGPE}) can be approximated by the GPE with an effective complex potential given by $\zeta(x) =\sum_{k} 2 |v_k(r)|^2-u_k(r)v_k^*(r)$ \cite{imag}. (Note the $y$ and $z$ dependencies manifest themselves only in the phases of the quantum amplitudes and thus cancel out when the phases cancel out).

To calculate the force arising from zero-point quantum fluctuations $F_x$, we must first solve the Bogoliubov equations for the quantum amplitudes.  To obtain a finite force we find it convenient to assume $\int^\infty_{-\infty} dx \Phi(x)=\tilde \Phi(0)=0$ where $\tilde \Phi(\lambda)$ is the potential in Fourier space. Extracting the trivial phase factor, i.e. $u_k(r)=U(x)e^{i {\bf k} \cdot {\bf r}}$ and $v_k(r)=V(x)e^{i {\bf k} \cdot {\bf r}}$, and working in Fourier space defined by $U(x)=\int d \lambda e^{i \lambda x} U(\lambda)$ and $V(x)=\int d \lambda e^{i \lambda x} V(\lambda)$, the Bogoliubov equations can be 
solved perturbatively to give the quasiparticle amplitudes to first order in $\eta$ as $U_1(k,\lambda)=\tilde{\Phi}(\lambda) \frac{\Gamma_U(k,\lambda)}{C(k,\lambda)} \mathsf{sign}(v_g^R)$ and 
$V_1(k,\lambda)=\tilde{\Phi}(\lambda) \frac{\Gamma_V(k,\lambda)}{C(k,\lambda)} \mathsf{sign}(v_g^R)$.
$\Gamma_U(k,\lambda)$ and $\Gamma_V(k,\lambda)$ are quantities easily derived from the Bogoliubov equations but, for the sake of clarity, we have chosen not to write out their full expressions as they would contribute little to the discussion.
%\begin{equation}
%\left(
%\begin{array}{cc}
%\Gamma_U(k,\lambda) \\
%\Gamma_V(k,\lambda)
%\end{array}
%\right)
%= 
%-\frac{1}{\lambda^2+\gamma^2}\left(
%\begin{array}{cc}
%a+c & b+d\\
%b-d & a-c\\
%\end{array}
%\right)
%\left(
%\begin{array}{cc}
%U_0(k) \\
%V_0 (k)
%\end{array}
%\right),
%\end{equation}
%, and the matrix elements are given by 
%$a=(-2 E_B+4 k^2 q k_x+4 q k_x)+(2 k^2 q+6 q+8 k_x^2 q) \lambda+(8 q  k_x+E_B) \lambda^2$, $b=(2 E_B-4 q kx)+(-6 q) \lambda$, $c=(-2 k^2+4 E_B q kx)+(-8 kx q^2+2 q E_B-4 kx) \lambda+(k^2-4 q^2-1) \lambda^2+2 kx \lambda^3+\lambda^4$, and $d=2 k^2+4 kx \lambda+3 \lambda^2$.
   The sign of the group velocity  of the reflected quantum fluctuation, denoted by $\mathsf{sign}(v^R_g)$,  arises from the boundary conditions where we exclude exponentially growing scattered waves and exclude incoming scattered waves due to causality. The group velocity of the reflected wave is given by $v_g^R=\frac{\partial E_k}{\partial \lambda_R}$ where the wavenumber of the reflected mode is given by $\lambda_R=k_x+k_R$ and $k_R$ is given by the non-trivial real root of the characteristic equation of the coupled Bogoliubov equations $C(k,\lambda)=\lambda[\lambda^3+ 4 k_x\lambda^2+ (4 k_x^2 +2 k^2+2 -2 q^2)\lambda+4 k_x(k^2+1)+2 \sqrt{2} qE_B]=0$. Assuming  $k_x=kf$, then $v_g^R>0$ if $-1<f<f_c$ and $v_g^R<0$ if $f_c<f<1$ where $f_c=-\frac{q \sqrt{k^2+2}}{\sqrt{2}(k^2+1)}$.

Next, these quasiparticle amplitudes can be used to determine the effective complex potential $\zeta(x)$ in the GGPE to give $\Psi^{(1)}(r)$.  Then, integrating over all momenta of the quantum fluctuations, the Casimir-like force due to the quantum fluctuations $F_x$ can be divided into two contributions as seen in eq. (\ref{fdiv}), given at the dominant order in $\eta$ as
\begin{equation}
\overline{F_{x}}=- \int d^3 k (F_{cond}(k)+ F_{fluc}(k))
\end{equation}
where  $\overline{F_{x}} = F_x/\eta^2 p_0 A \sqrt{\rho_0 a_{sc}^3}$ and the zeroth order interaction pressure is given by $p_0=g \rho_0^2/2$.  The contribution to the force due to the condensate modified by the quantum fluctuations is given by
\begin{eqnarray}
&& F_{cond}(k) = \mathsf{Res}_{\lambda=k_R} \hspace{0.5 mm} \frac{8 \sqrt{2}}{ \sqrt{{\pi}} }
\tilde{\Phi}(\lambda)
 \lambda/(\lambda^2+2-2q^2) \nonumber \\
&& \Big\{ [U_0(k)(1+\frac{\sqrt{2}q}{\lambda})-4 V_0(k)] \tilde
V_1(\lambda,k)+V_0(k)(1-\frac{\sqrt{2} q}{\lambda}) \tilde U_1(\lambda,k) \Big\} 
\end{eqnarray}
and the contribution to the force given directly by the quantum fluctuations is given by 
\begin{equation}
F_{fluc}(k) =\mathsf{Res}_{\lambda=k_R} \hspace{0.5 mm} \frac{8 \sqrt{2}}{ \sqrt{{\pi}} } 
\tilde{\Phi}(\lambda) V_0(k) \lambda \tilde V_1(\lambda,k),
\end{equation}
where $\mathsf{Res}_{\lambda=k_R}$ is the residue at $\lambda=k_R$.  The zeroth order quantum amplitudes (for a homogeneous gas at rest) are given by $U_0(k)=\sqrt{\frac{1}{2} \left(\frac{k^2+1}{E_B}+1 \right)}$ and $V_0(k)=\sqrt{\frac{1}{2}
  \left( \frac{k^2+1}{E_B}-1 \right)}$.

Finally, to illustrate the calculation of $F_x$, we shall define a specific potential describing the stationary object in the flow as $\tilde{\Lambda}(\lambda) \equiv [\tilde{\Phi}(\lambda)]^2= \frac{1}{\Delta}$ for $|\lambda + k_0| < \Delta/2$, $|\lambda - k_0| < \Delta/2$ and zero otherwise.  In real space this can be written as $\Lambda(x)=\frac{\sin(x \Delta/2)}{x \pi \Delta} 2 \cos(k_0 x) $ where $\Delta$ is positive, $1/\Delta$ is a measure of the width of the potential in real space and $k_0$ is a measure of the typical wavenumber of the potential in real space. Both $\Delta$ and $k_0$ are normalized by the healing length.  We assume $k_0>\Delta/2$ so that the potential satisfies the condition $\tilde \Phi(0)=0$.  Note that as $\Delta \rightarrow 0$ the potential becomes periodic in real space and delocalized; our analysis would then no longer apply.  $\overline{F_{x}}$ peaks when the width in Fourier space (or real space) is on the order of the healing length, i.e. $\Delta \approx 1$ and disappears in the localized and delocalized limits.

\begin{figure}[]
%\begin{center}
%\includegraphics[scale=0.3,angle=270]{test.eps}
\includegraphics[scale=0.7]{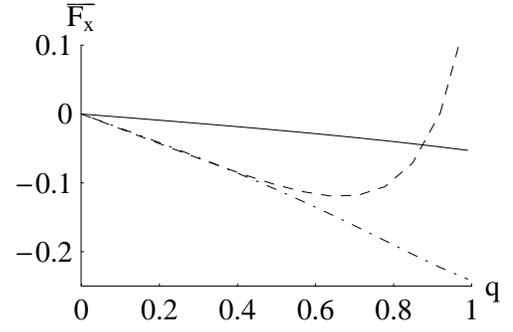}
%\end{center}
\caption{\label{forceplot}The scaled force due to quantum fluctuations, $\overline{F_{x}} = F_x/\eta^2 p_0 A \sqrt{\rho_0 a_{sc}^3}$, acting on a stationary localized potential described by the parameters $k_0$ and $\Delta$ as a function of the mach number of the flow, $q$ (measured far from the potential, i.e. at $x=\infty$), for three different values of $k_0$: $k_0=5.0$ is given by the 
solid line, $k_0=2.4$ is given by the dashed-dotted line, $k_0=1.8$ is given by the dashed line.  The width of the potential in real (and Fourier) space is equal to the healing length, i.e. $\Delta=1$.  ($k_0$ and $\Delta$ are both normalized by the healing length.)}
\end{figure}

In the rest frame of $\Phi(r)$, the grand canonical energy decreases with increasing flow speed.  It follows from this that the flow would accelerate in the presence of dissipation, implying a negative drag force similar to the behavior of a moving gray soliton \cite{gsoliton}. At larger $k_0$ such behavior is exhibited by $\overline{F_{x}}$ as seen in Figure \ref{forceplot}.   At smaller $k_0$ or, equivalently, at larger characteristic wavelengths of the potential in real space,  an instability arises as the flow speed increases.  This instability is consistent with the dynamical modulation instability observed for a perfect lattice in the GPE \cite{dinstability}, which also occurs at long characteristic wavelengths (small $k_0$) and large speeds.  It is also perhaps  instructive to recall the non-trivial and highly geometry-dependent sign of the Casimir force in the EM vacuum (for example, a set-up of parallel conductors in an EM vacuum leads to an attractive force while a cubical cavity leads to a repulsive force).

%The force arising from the condensate modified from the fluctuations, 
%$F_{cond}$ can be determined by the GGPE eq. \ref{}, which describes 
%$\Psi_1$ to leading order
%\begin{equation}
%\end{equation}
%where $f$ insures orthogonality between the condensate and the excited modes 
%but does not contribute to the force to this order of the calculation.  
%Since $\eta\ll1$ the GGPE can be approximated as the GPE with a complex 
%potential
%\begin{equation}
%\end{equation}

Even though this Casimir-like force exists for speeds much lower than the speed of sound, which in  
Bogoliubov's theory is equal to Landau's critical velocity for the onset of dissipation,  
this drag force does not explicitly violate the spirit of Landau's principle \cite{pomeau}.  Let us recall that Landau's principle  
states that above a critical velocity a system can lower its energy by the creation of  
quasiparticles \cite{landau}.  In our analysis, however, we do not assume that any quasiparticles are created, i.e. the  
system remains at $T=0$ in the sense defined above. 
This drag force arises from the scattering of zero-temperature quantum fluctuations and is not caused by  
the creation of quasiparticles, which must satisfy the Landau criterion; instead it is  
caused by the changing nature of the eigenstates of the quantum fluctuations, akin to the  
original Casimir force between two conductors.  

In this letter, we have shown in a specific example that a finite Casimir-like drag force arises in a moving BEC at $T=0$.  In this case, unlike for the nucleation of vortices, there is no free energy barrier to cross and, at least for this particular situation, the effective critical velocity is zero.  Since a nonzero effective critical velocity does exist 
at the dominant order (on the order of the mean field), one would expect to find the semblance of a nonzero critical velocity as seen in \cite{macroscopic}, even though, at least in the case considered here, the actual critical velocity for the system might be zero.

We expect this Casimir-like force to act upon any density perturbation, including those created by laser fields \cite{macroscopic}, untrapped impurities \cite{impurity}, vortices, etc.,  moving relative to the condensate.  We also expect this force to be more apparent in condensates of lower dimensions due to the enhancement of quantum fluctuations.  Finally, although the present analysis assumes a dilute gas and does not strictly apply to dense systems, we do expect a Casimir-like force from quantum fluctuations also to exist for superfluid liquid Helium.  In fact, the Casimir-like force might have a stronger effect in liquid Helium since quantum fluctuations dominate the Helium condensate at $T=0$.

We conclude by noting that although we have discussed a force that exists on a stationary object in a superfluid moving at any arbitrarily small velocity, these results are not inconsistent with the existence of persistent superfluid currents in toroidal geometries.  In this letter where we consider an infinite medium, the Casimir-like force arises from the nonlocal perturbation of the scattered quantum fluctuations.  These scattered fluctuations can be seen to transport energy far from the potential similar to wave-drag situations in classical fluids.  However, in a finite geometry such as a superflow in a torus the scattered waves will interact with the localized object.  In the steady state, we expect these backscattered waves to eventually cancel out the effect discussed in this paper and thus remain consistent with a persistent superflow. In this system, the experimental manifestation of this Casimir-like effect would not be a drag force at arbitrarily low speeds but rather the presence of these scattered waves which could, in principle, be observed.  

The authors would like to thank I. Carusotto, K. Burnett, C. Connaughton, D. Gu\'ery-Odelin, A. Fetter, S. Morgan, V. Hakim, and S. Stringari for valuable discussions.


\begin{thebibliography}{9}
\bibitem{emdrag} This is no longer the case, 
however, at
nonzero temperature \cite{casimirdrag} or in a non-inertial frame, e.g. a 
rotating dielectric
should experience a resistive Casimir torque \cite{torque}.
\bibitem{emcons}  Due to phase and time translation symmetry, mass and 
energy are conserved in our system, and
the usual picture of an 'irreversible' phenomenon with dissipation in 
the ordinary sense does not apply.
\bibitem{mac} We assume that the $k=0$ mode is macroscopically occupied in the rest frame of the condensate, consistent with the Bogoliubov approach \cite{bog,fetter}.
\bibitem{modv} Note that for the specific potential considered below, $\int^\infty_{-\infty} dx \Phi(x)=\tilde \Phi(0)=0$. This critical flow velocity in our nonuniform case is equal to the critical velocity for the uniform case (i.e. the speed of sound) to the order of the calculation \cite{Hakim}.
\bibitem{scatter} Specifically, we require that the spatial-temporal correlation function - formed by the expectation value of the incoming fluctuation operators taken with respect to this quantum state - be identical to that of a Galilean-transformed uniform ground state.  Furthermore, this condition affects only the incoming waves as the correlation functions are defined in such a way that the spatial-temporal difference describes a wave that propagates towards the stationary object.
\bibitem{imag} This effective complex potential implies that there is mass transfer between the condensate (described by wave function $\Psi^{(1)}$) and the quantum fluctuations.
\end{thebibliography}
\end{document}